\documentclass[sigplan, nonacm]{acmart}
 
\renewcommand{\shortauthors}{} 

\usepackage{colortbl}   
\usepackage{url}
\usepackage{graphicx}
\usepackage{subcaption}
\usepackage{pifont}
\usepackage{wrapfig}

\usepackage{tabularx}
\usepackage{booktabs}
\usepackage{multirow}

\usepackage{amsmath}
\usepackage{makecell}

\usepackage{algorithm}
\usepackage{algpseudocode}

\AtBeginDocument{%
  }

\setcopyright{acmlicensed}
\copyrightyear{2018}
\acmYear{2018}
\acmDOI{XXXXXXX.XXXXXXX}
\acmConference[Conference acronym 'XX]{Make sure to enter the correct
  conference title from your rights confirmation email}{June 03--05,
  2018}{Woodstock, NY}
\acmISBN{978-1-4503-XXXX-X/2018/06}

\begin{document}

\title[ ]{Accelerating 3D Gaussian Splatting using Tensor Cores}  


\author{Sheng Li}
\affiliation{%
  \institution{University of Pittsburgh}
  \country{}
}
\email{shl188@pitt.edu}

\author{Yang Sui}
\affiliation{%
  \institution{Rice University}
  \country{}
  }
\email{yangsui1994@gmail.com}

\author{Yue Wu}
\affiliation{%
  \institution{New York University}
  \country{}
}
\email{yw4927@nyu.edu}

\author{Zhuoran Song}
\affiliation{%
  \institution{Shanghai Jiao Tong University}
  \country{}
}
\email{songzhuoran@sjtu.edu.cn}

\author{Bo Yuan}
\affiliation{%
  \institution{Rutgers University}
  \country{}
  }
\email{bo.yuan@soe.rutgers.edu}

\author{Xulong Tang}
\affiliation{%
  \institution{University of Pittsburgh}
  \country{}
  }
\email{tax6@pitt.edu}

\author{Yue Dai}
\affiliation{%
  \institution{Illinois Institute of Technology}
  \country{}
  }
\email{ydai21@illinoistech.edu}

\renewcommand{\shortauthors}{Li et al.}


\newcommand{\GY}[1]{\textcolor{blue}{Geng: #1}}
\newcommand{\todo}[1]{\textcolor{red}{\sf\bfseries Todo: #1}}
\newcommand{\bred}[1]{\textcolor{red}{\sf\bfseries #1}}
\newcommand{\red}[1]{\textcolor{red}{#1}}
\newcommand{\blue}[1]{\textcolor{blue}{#1}}
\newcommand{\yellow}[1]{\textcolor{yellow}{#1}}
\newcommand{\purple}[1]{\textcolor{purple}{#1}}
\newcommand{\brown}[1]{\textcolor{brown}{#1}}
\newcommand{\cross}[1]{\textcolor{red}{\sout{#1}}}
\newcommand{\SL}[1]{\textcolor{purple}{Sheng: #1}}
\newcommand{\YD}[1]{\textcolor{teal}{Yue: #1}}

\newcommand{\xulong}[1]{\COMMENT{\textcolor{cyan}{\sf\bfseries Xulong: #1}}}
\newcommand{\Yue}[1]{\textcolor{teal}{\textbf{Yue: #1}}}

\newcommand{\squishlist}{
    \begin{list}{$\bullet$}
        { \setlength{\itemsep}{0pt}      \setlength{\parsep}{0pt}
            \setlength{\topsep}{0.5pt}       \setlength{\partopsep}{0pt}
            \setlength{\listparindent}{-2pt}
            \setlength{\itemindent}{-5pt}
            \setlength{\leftmargin}{0.5em} \setlength{\labelwidth}{0em}
            \setlength{\labelsep}{0.2em} } }
    
\newcommand{\squishend}{
\end{list}  }

\newcommand{\yw}[1]{\textcolor{blue}{[yw:~#1]}}

\begin{abstract}
3D Gaussian Splatting (3DGS) has become a leading technique for real-time neural rendering and 3D scene reconstruction, but its rendering cost remains too high for many latency-sensitive scenarios. 
In particular, the rasterization stage in 3DGS dominates end-to-end rendering time, during which the renderer repeatedly evaluates each Gaussian's contribution to each covered pixel, making this stage compute-bound. 
At the same time, modern GPUs provide high-throughput Tensor Cores for low-precision matrix operations, yet existing 3DGS systems execute rasterization entirely on CUDA cores and leave Tensor Cores idle. We find that 3DGS rendering can be executed in FP16 with negligible quality degradation, suggesting a promising opportunity for Tensor Core acceleration.
However, exploiting Tensor Cores for 3DGS is non-trivial because rasterization does not naturally match their execution model. Existing 3DGS rasterization is expressed as irregular per-pixel scalar operations, whereas Tensor Cores require dense, regular, and reuse-rich matrix workloads. Moreover, conventional tile-by-tile execution fails to exploit Gaussian reuse across neighboring tiles, resulting in repeated data loading and thus high data movement overhead.
To this end, we present TensorGS, a 3DGS acceleration framework using Tensor Cores. TensorGS tensorizes the dominant rasterization computation into Tensor-Core-compatible matrix operations and introduces cross-tile grouping to improve Gaussian reuse, amortize overhead, and increase Tensor Core utilization. Experimental results show that TensorGS improves end-to-end rendering performance by 1.65$\times$ while preserving image quality.
\end{abstract}


\maketitle

\section{Introduction}
\label{sec:intro}

3D Gaussian Splatting (3DGS) has emerged as a leading representation for high-quality real-time neural rendering and 3D scene reconstruction~\cite{10.1145/3592433, kim20243dgs, qiu2025advancing, reddy2025survey}. 
Compared with prior radiance-field approaches such as NeRF~\cite{mildenhall2021nerf}, 3DGS represents a scene explicitly as a set of anisotropic Gaussians and renders them via rasterization rather than dense neural-network evaluation, enabling substantially faster rendering while preserving high visual quality~\cite{10.1145/3592433, hollein20253dgs, xu2025gaussian}. 
This explicit representation makes 3DGS particularly attractive in practice, since it offers a favorable balance between rendering efficiency and visual quality.
As a result, 3DGS has quickly become an important foundation for interactive view synthesis and real-time 3D scene rendering. 
Due to these advantages, 3DGS has been increasingly adopted in latency-sensitive applications such as virtual and augmented reality, robotics, autonomous systems, and interactive digital-twin platforms~\cite{zhu20243d, liao20253dgs, matias2025volume, zhou2025visual}.

Despite its efficiency advantages over NeRF-style methods, 3DGS rendering remains too expensive for many latency-sensitive scenarios, especially on large scenes and high-resolution outputs. 
A standard 3DGS pipeline consists of frustum culling, feature computation, Gaussian sorting, and rasterization~\cite{10.1145/3592433}. 
Among these stages, rasterization dominates end-to-end rendering time. Within rasterization, the renderer repeatedly evaluates each Gaussian's contribution to each covered pixel, including a quadratic power computation followed by opacity calculation and alpha blending. 
This per-pixel, per-Gaussian arithmetic scales rapidly with both scene complexity and image resolution, making rasterization the performance bottleneck in 3DGS.
Our profiling shows that rasterization accounts for about 75\% of end-to-end rendering time on average, while the power computation alone contributes 68\% of rasterization time. 
Consequently, a substantial latency gap remains between current 3DGS systems and the requirements of real-time applications.
For example, real-time VR typically requires a per-frame latency below \textit{16}~ms~\cite{noh2025flexnerfer}. Although the exact rendering cost of 3DGS depends on the scene, the standard 3DGS implementation~\cite{10.1145/3592433} can still take around \textit{30}~ms to render an image at 1K resolution on several representative scenes.
Hence, it is urgent to develop a faster 3DGS framework to close the gap.

At the same time, modern GPUs provide a large pool of specialized compute resources that existing 3DGS systems fail to exploit. 
In particular, NVIDIA Tensor Cores deliver substantially higher throughput than conventional CUDA cores for supported low-precision matrix operations~\cite{markidis2018nvidia}. 
However, today's 3DGS implementations execute rasterization entirely on CUDA cores, leaving Tensor Cores essentially idle throughout the rendering pipeline. 
Our measurements reveal a striking mismatch: while rasterization already exhibits high streaming multiprocessor activity (\textit{77}\% SM utilization), Tensor Core utilization remains zero. This suggests that current 3DGS pipelines are compute-bound only with respect to the subset of GPU arithmetic units they actively use, while a much larger pool of matrix-specialized hardware remains underutilized. 
Meanwhile, our empirical study shows that 3DGS rendering can be executed in FP16 with negligible quality degradation, indicating that low-precision Tensor Core execution is feasible. 
Together, these results suggest a promising opportunity to accelerate 3DGS rasterization with Tensor Cores.

However, 3DGS rasterization does not naturally match the execution model of Tensor Cores for two reasons.
First, Tensor Cores are designed for dense matrix-style operations, whereas 3DGS rasterization is originally expressed as irregular per-pixel scalar computation. This mismatch requires reorganizing per-pixel Gaussian interactions into matrix operands with Tensor-Core-compatible shapes and layouts while keeping the transformation overhead low.
Second, Tensor Cores achieve high throughput only when supplied with sufficiently large, regular, and reuse-rich workloads, but 3DGS rasterization is organized around small per-tile workloads with limited regularity. 
Although neighboring tiles often share overlapping Gaussians, conventional tile-by-tile execution fails to exploit this cross-tile reuse, leading to repeated Gaussian loading and poor Tensor Core utilization. 
Therefore, simply tensorizing the original computation is insufficient to fully unlock the performance potential of Tensor Cores.

To address the challenges, we present \textbf{TensorGS}, a 3D Gaussian Splatting acceleration framework that enables practical Tensor Core execution for rasterization on GPUs. 
TensorGS is built on two key ideas. 
First, it \textit{tensorizes the dominant power computation} by reformulating batches of Gaussian--pixel interactions into Tensor-Core-compatible matrix operations, allowing the core rasterization arithmetic to execute through WMMA instructions rather than scalar CUDA operations. 
Second, TensorGS introduces \textit{cross-tile grouping} to improve Gaussian reuse across neighboring tiles, amortize data-preparation overhead, and make Tensor Core execution more efficient. Together, these two techniques bridge the gap between the irregular structure of 3DGS rasterization and the execution requirements of Tensor Cores.
Our contributions are summarized as follows:
\begin{itemize}
    \item We identify that the dominant bottleneck in 3DGS rasterization is compatible with Tensor Core acceleration, and show that low-precision execution preserves rendering quality.
    \item We propose TensorGS, a Tensor-Core-aware 3DGS rasterization framework that combines computation tensorization with cross-tile grouping to improve both hardware compatibility and data reuse.
    \item Extensive experiments show that our proposed TensorGS framework effectively accelerates 3DGS rendering by delivering 1.65$\times$ speedup while maintaining image quality.
\end{itemize}

\section{Background}
\label{sec:bg}

3D Gaussian Splatting (3DGS) is an explicit radiance field representation that models a scene as a set of learnable anisotropic 3D Gaussians rather than an implicit neural field~\cite{10.1145/3592433}. Each Gaussian is parameterized by a 3D mean, a covariance matrix, an opacity, and spherical harmonics (SH) coefficients. The mean specifies the Gaussian center in 3D space, while the covariance matrix controls its spatial extent, shape, and orientation. The opacity determines the strength of its contribution during rendering, and the SH coefficients represent its view-dependent color as a function of the viewing direction. For clarity, Table~\ref{tab:bg_notation} summarizes the core notation used throughout the paper. 

In practice, the Gaussians are typically initialized from a sparse Structure-from-Motion (SfM) point cloud and then optimized through differentiable rendering. Compared with NeRF-style methods that require evaluating a neural network at many sampled points along each ray~\cite{mildenhall2021nerf}, 3DGS uses explicit rasterization primitives and therefore enables substantially more efficient rendering while maintaining high-quality novel view synthesis.
A typical 3DGS rendering pipeline consists of four major stages: \emph{frustum culling}, \emph{feature computation}, \emph{Gaussian sorting}, and \emph{rasterization}.
The illustration of these four stages is depicted in Figure \ref{fig:3DGS_TensorGS_overview}.

\begin{table}[t]
    \centering
    \caption{Core notation used in the paper.}
    \label{tab:bg_notation}
    \small
    \begin{tabular}{l p{0.77\linewidth}}
        \toprule
        Symbol & Meaning \\
        \midrule
        $\boldsymbol{\mu}_i$ & 3D mean of the $i$-th Gaussian \\
        $\boldsymbol{\Sigma}_i$ & 3D covariance matrix of the $i$-th Gaussian \\
        $\boldsymbol{\mu}'_i$ & Projected 2D mean of the $i$-th Gaussian on the image plane \\
        $\boldsymbol{\Sigma}'_i$ & Projected 2D covariance matrix of the $i$-th Gaussian on the image plane \\
        $o_i$ & Opacity of the $i$-th Gaussian \\
        $\mathbf{c}_i$ & Color of the $i$-th Gaussian \\
        $\mathbf{x}$ & Pixel location on the image plane \\
        $\alpha_i$ & Opacity contribution of the $i$-th Gaussian at pixel $\mathbf{x}$ \\
        $T_i$ & Transmittance before blending the $i$-th Gaussian \\
        \bottomrule
    \end{tabular}
\end{table}

\textbf{Frustum culling (Pre-process).}
Given a camera viewpoint, the renderer first removes Gaussians that are invisible to the current view. This step discards primitives (i.e., Gaussians) outside the viewing frustum before more expensive screen-space processing begins and reduces the number of Gaussians that must be processed in later stages.

\textbf{Feature computation.}
For each visible Gaussian, 3DGS projects its 3D mean and covariance from world space to the image plane, 
producing a 2D mean $\boldsymbol{\mu}'_i$ and a 2D covariance matrix $\boldsymbol{\Sigma}'_i$ that define the position, size, shape, and orientation of the Gaussian's elliptical splat in screen space.
At the same time, the renderer evaluates the Gaussian's view-dependent color $\mathbf{c}_i$ using its SH coefficients and the current viewing direction. After projection, the image plane is partitioned into a regular grid of tiles, typically of size $16\times16$ pixels. The renderer then performs a tile--Gaussian intersection test to determine which tiles are overlapped by each projected Gaussian. In practical implementations, an axis-aligned bounding box (AABB) of the projected Gaussian is often used for this test. For every overlapped tile, a duplicated Gaussian entry is emitted. Therefore, if one Gaussian overlaps $k$ tiles, it produces $k$ duplicated entries. These duplicated entries preserve tile-local parallelism in later stages, but they also increase the overhead of sorting and rasterization.

\textbf{Gaussian sorting.}
Because 3DGS uses front-to-back alpha blending, the Gaussian entries associated with the same tile must be sorted by depth before rendering. After tile assignment, the duplicated entries belonging to each tile are grouped together and sorted according to their depths, typically using a parallel sorting algorithm such as radix sort on GPUs. This stage establishes the correct visibility order for Gaussian rendering.

\textbf{Rasterization (Volume Rendering).}
Finally, the sorted Gaussian entries are blended to produce the output image. For a pixel location $\mathbf{x}$ and the $i$-th Gaussian in the depth-sorted list of a tile, let
\begin{equation}
\label{eq:power_d}
    \mathbf{d}_i = \mathbf{x} - \boldsymbol{\mu}'_i
\end{equation}
denote the offset from the pixel to the projected Gaussian center. The renderer then evaluates the quadratic power term
\begin{equation}
\label{eq:power}
    \mathrm{power}_i = -\frac{1}{2}\mathbf{d}_i^\top (\boldsymbol{\Sigma}'_i)^{-1}\mathbf{d}_i,
\end{equation}
and converts it to opacity
\begin{equation}
    \alpha_i = o_i \exp(\mathrm{power}_i).
\end{equation}
Using front-to-back alpha blending, the final pixel color is accumulated as
\begin{equation}
    \mathbf{C} = \sum_{i=1}^{N} T_i \alpha_i \mathbf{c}_i,
    \qquad
    T_i = \prod_{j=1}^{i-1}(1-\alpha_j),
\end{equation}
where $T_i$ denotes the transmittance before processing the $i$-th Gaussian.

In practical GPU implementations, rasterization is usually performed in a tile-based manner. The image is divided into small tiles, and each tile is assigned to one thread block. Within a thread block, one thread is typically responsible for one pixel, so all pixels in the tile iterate over the same depth-sorted Gaussian list in parallel. To reduce unnecessary work, implementations commonly prune negligible Gaussian contributions when their opacity is sufficiently small and terminate processing early once the remaining transmittance becomes negligible. Throughout this paper, we refer to the evaluation of $\mathrm{power}_i$ as the \emph{power computation}, since it is the dominant arithmetic operation in rasterization.

\section{3DGS Pipeline Analysis and Motivation}
\label{sec:moti}

\subsection{Rasterization Bottleneck and Underutilized Tensor Cores}
\label{sec:moti_underutilized}



We begin by characterizing where time is spent in the 3DGS rendering pipeline and how the underlying GPU resources are utilized. 

\textbf{Experimental setup.}
Unless otherwise stated, all profiling results in this section are collected on a server equipped with an NVIDIA A100 GPU and the official 3DGS implementation~\cite{10.1145/3592433}. We evaluate six representative scenes from four widely used datasets: Tanks\&Temples~\cite{knapitsch2017tanks}, Deep Blending~\cite{hedman2018deep}, Synthetic-NeRF~\cite{mildenhall2021nerf}, and Synthetic-NSVF~\cite{liu2020neural}. Kernel-level metrics (SM utilization, Tensor Core pipe utilization, DRAM bandwidth, etc.) are obtained with NVIDIA Nsight Compute, and end-to-end timing is measured with CUDA events averaged over 100 rendered frames per scene.

\begin{figure}[t]
    \centering
    \includegraphics[width=1\linewidth]{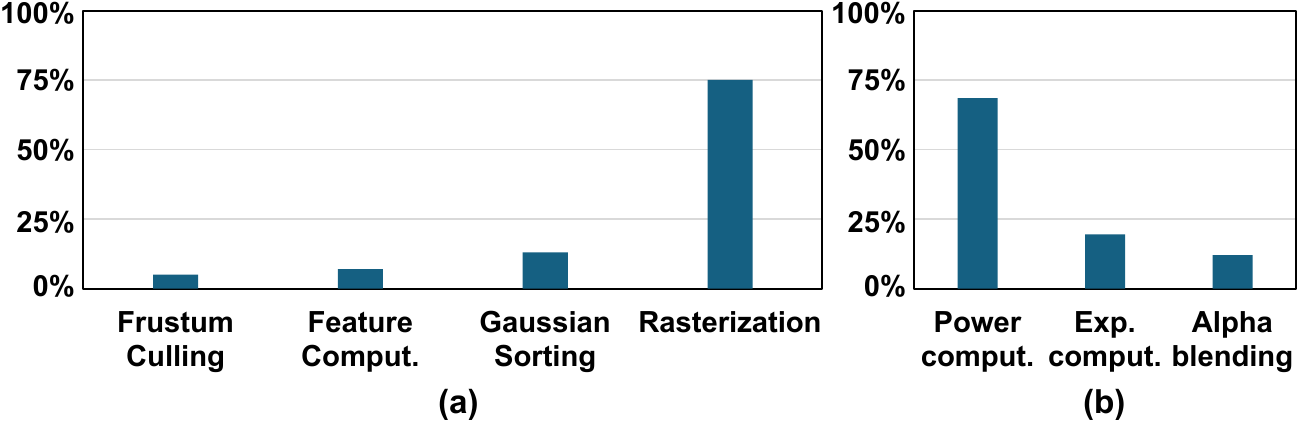}
    \caption{(a) Averaged end-to-end time breakdown of the 3DGS rendering pipeline across six representative scenes. (b) Averaged time breakdown within the rasterization stage.}
    \label{fig:moti_pipeline_breakdown}
    \label{fig:moti_raster_breakdown}
\end{figure}

\textbf{Rasterization dominates the rendering pipeline.} 
As described in Section~\ref{sec:bg}, a typical 3DGS rendering pipeline consists of four stages: \textit{frustum culling}, \textit{feature computation}, \textit{gaussian sorting}, and \textit{rasterization}.
Figure~\ref{fig:moti_pipeline_breakdown}(a) presents the averaged time breakdown of these four stages across the six representative scenes. 
As shown in the figure, rasterization accounts for about 75\% of the end-to-end rendering time, making it the largest contributor and naturally the optimization target.

As mentioned in Section~\ref{sec:bg}, rasterization consists of several per-pixel operations. For each Gaussian--pixel pair, the renderer first evaluates the Gaussian contribution by computing the power term $(\mathrm{power}_i = -\frac{1}{2}\mathbf{d}_i^\top (\boldsymbol{\Sigma}'_i)^{-1}\mathbf{d}_i),$
and then converting it to opacity $(\alpha_i = o_i \exp(\mathrm{power}_i))$ with a exponential computation.
The renderer then prunes negligible contributions when the resulting opacity is sufficiently small.
After that, it performs alpha blending and transmittance update for the surviving contributions, and finally applies early termination when the remaining transmittance becomes negligible. 
Figure~\ref{fig:moti_raster_breakdown}(b) reports the time breakdown of these operations. Among them, the power computation alone consumes 68\% of rasterization time, corresponding to 51\% of the end-to-end time of 3DGS. This result identifies the power computation as the dominant hotspot.


\begin{table}[t]
    \centering
    \caption{Hardware utilization during Rasterization stage on an NVIDIA A100 GPU, measured with NVIDIA Nsight Compute. Rasterization exhibits high SM activity, but Tensor Cores remain completely unused.}
    \label{tab:moti_hw_util}
    \begin{tabular}{lc}
        \toprule
        Metric & 3DGS Rasterization \\
        \midrule
        SM utilization                     & 77\% \\
        DRAM bandwidth utilization         & 38\% \\
        \textbf{Tensor Core utilization}   & \textbf{0\%} \\
        \bottomrule
    \end{tabular}
\end{table}

\begin{table*}[t]
    \centering
    \caption{Rendering quality under FP32 and FP16 execution across six representative scenes. All metrics are computed against the ground-truth images from the corresponding test views. Here, $\Delta$ denotes the difference between the FP16 result and the FP32 baseline.}
    \label{tab:moti_fp16_quality}
    \begin{tabular}{p{2.8cm} p{2cm} m{1.35cm}<{\centering} m{1.35cm}<{\centering} m{1.35cm}<{\centering}
        m{1.35cm}<{\centering} m{1.35cm}<{\centering} m{1.35cm}<{\centering}}
        \toprule
        \multirow{2}{*}{Dataset} & \multirow{2}{*}{Scene}
        & \multicolumn{3}{c}{PSNR $\uparrow$}
        & \multicolumn{3}{c}{LPIPS $\downarrow$} \\
        \cmidrule(lr){3-5}
        \cmidrule(lr){6-8}
        & & FP32 & FP16 & $\Delta$
          & FP32 & FP16 & $\Delta$ \\
        \midrule
        \multirow{2}{*}{Tanks\&Temples} & Train     & 24.17 & 24.22 & +0.05 & 0.173 & 0.175 & +0.002 \\
                                        & Truck     & 26.46 & 26.49 & +0.03 & 0.224 & 0.223 & -0.001 \\
       \multirow{2}{*}{Deep Blending}   & Playroom  & 29.89 & 29.82 & -0.07 & 0.218 & 0.215 & -0.003 \\
                                        & DrJohnson & 35.19 & 35.10 & -0.09 & 0.179 & 0.181 & +0.002 \\
        Synthetic-NeRF & Lego           & 34.47 & 34.40 & -0.07 & 0.022 & 0.020 & -0.002 \\
        Synthetic-NSVF & Palace         & 33.75 & 33.78 & +0.03 & 0.030 & 0.031 & +0.001 \\
        \bottomrule
    \end{tabular}
\end{table*}

\textbf{Tensor Cores are completely idle.}
The above breakdown suggests that rasterization is compute-bound, consistent with our hardware-level measurements: as shown in Table~\ref{tab:moti_hw_util}, the SM utilization during rasterization reaches 77\% on an NVIDIA A100 GPU, while the DRAM bandwidth utilization is below 40\%. 
Hence, there is little room for further acceleration on general-purpose CUDA cores.

However, a closer look at the GPU's full compute hierarchy reveals a substantially different picture. Modern NVIDIA GPUs expose two independent classes of arithmetic units on every SM: the general-purpose CUDA cores and the matrix-specialized Tensor Cores~\cite{choquette2021nvidia}. On an NVIDIA A100, the Tensor Cores deliver up to 312~TFLOPS of FP16 throughput, which is over 16$\times$ the FP32 throughput of the CUDA cores (19.5 TFLOPS). 
As Table \ref{tab:moti_hw_util} shows, the Tensor Core utilization during 3DGS rasterization is essentially 0\%: not a single matrix instruction is issued throughout the entire rendering pipeline. In other words, the kernel appears compute-bound only because it is constrained to a small fraction of the GPU's compute resources; the vast majority sits completely idle.
In the following subsections, we conduct preliminary experiments to explore whether we can leverage these idle Tensor Cores to accelerate the 3DGS pipeline.

\subsection{Opportunity and Challenges}
\label{sec:moti_tc}

\subsubsection{Low-Precision Feasibility}
\label{sec:moti_low}
\hfill\break 
To benefit from Tensor Cores, the target computation must be executable in a Tensor-Core-friendly precision~\cite{choquette2021nvidia}, such as FP16, without compromising numerical stability.
Considering that the typical 3DGS pipeline is executed in FP32 precision, a second question arises for 3DGS rasterization: can the 3DGS pipeline be executed in lower-precision FP16 without degrading rendered image quality? 

To answer this question, we compare the standard FP32 implementation with an FP16 version across six representative scenes. Specifically, for each scene, we train two versions of the 3DGS model, one in FP32 precision and the other in FP16 precision. Then, for the same test views, we render each scene in both FP32 and FP16, and evaluate the rendered image quality against the corresponding ground-truth images using two standard metrics: PSNR and LPIPS. 
As shown in Table~\ref{tab:moti_fp16_quality}, the image quality is preserved: the average PSNR and average LPIPS remain essentially unchanged. These results indicate that FP16 execution preserves rendering quality to a visually identical level, making Tensor-Core acceleration practical for 3DGS rasterization.

\subsubsection{Naive Tensorization Is Feasible but Not Enough}
\label{sec:moti_naive}
\hfill \break
Importantly, idle Tensor Cores can be exploited only when the dominant workload can be expressed in a matrix multiplication form. This is indeed the case for the power computation in 3DGS rasterization. As shown in Equations~\ref{eq:power_d} and~\ref{eq:power}, for a tile of pixels and a batch of Gaussians, the same quadratic power term must be calculated repeatedly for every (Gaussian, pixel) pair. Although the baseline implementation performs these calculations in a scalar thread-per-pixel manner on CUDA cores, the computation itself can be reorganized across the Gaussian and pixel dimensions into a dense matrix form. This observation shows that a straightforward tensorization of the power computation is feasible in principle. The detailed tensor reformulation of our design is presented in Section \ref{sec:design_tensorizing}.

However, migrating the computation to Tensor Cores does not automatically guarantee meaningful end-to-end speedups. 
In our practice, a naive tensorization of the power computation still suffers from low Tensor Core utilization (details are presented in Section \ref{sec:design_limitation}).
After the power computation is migrated to Tensor Cores, the rasterization pipeline becomes increasingly constrained by data movement, and Gaussian loading emerges as a major limitation. 
Therefore, effectively exploiting Tensor Cores for 3DGS requires not only a tensor reformulation of the hotspot computation, but also a careful execution design that sustains high Tensor-Core efficiency. 

\section{TensorGS Design}
\label{sec:design}

\subsection{Design Overview}
\label{sec:design_overview}

\begin{figure*}[t]
    \centering
    \includegraphics[width=0.95\textwidth]{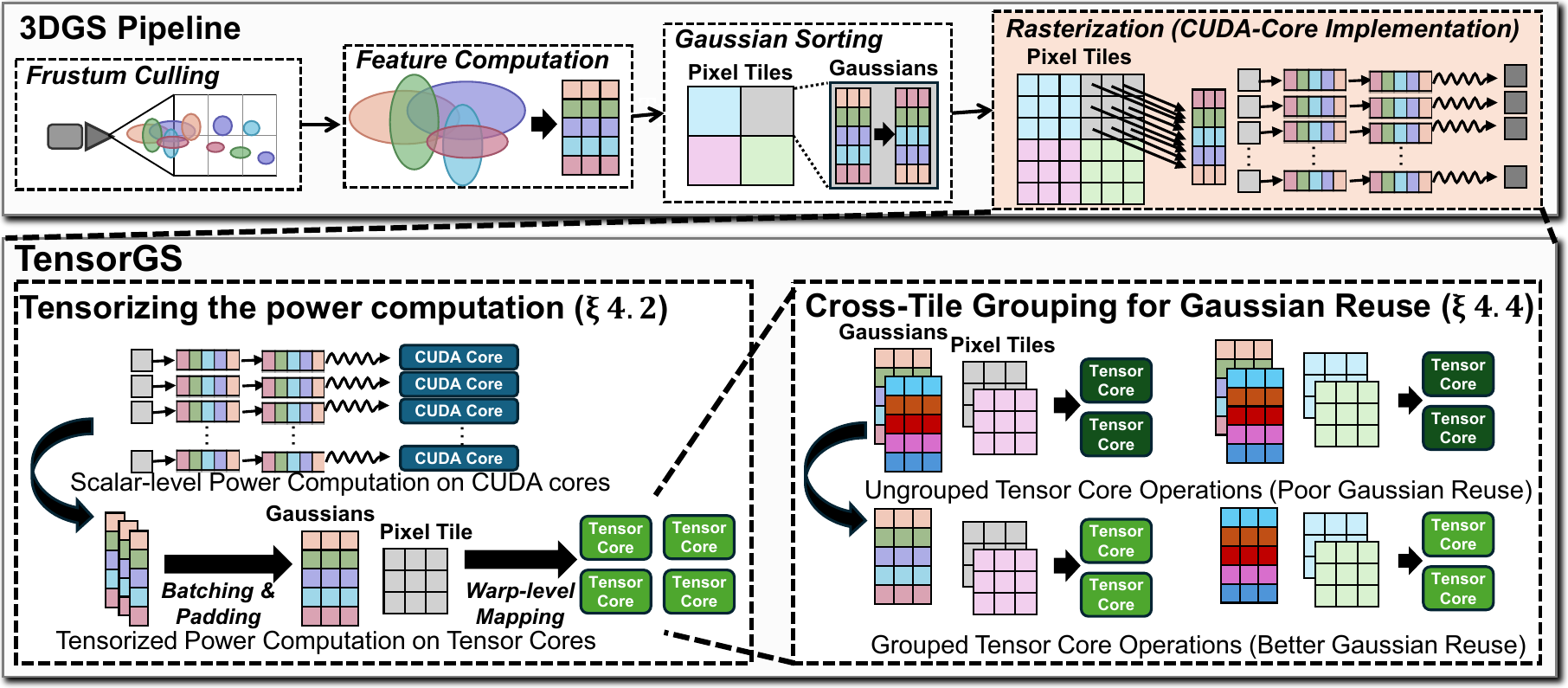}
    \caption{Overview of the original 3DGS pipeline and our proposed TensorGS.}
    \label{fig:3DGS_TensorGS_overview}
\end{figure*}

In this section, we present \textbf{TensorGS}, our Tensor-Core accelerated 3DGS rasterization design. 
The goal of TensorGS is to reorganize this hotspot to better match the execution model of modern NVIDIA Tensor Cores.
TensorGS consists of two key components, as shown in Figure \ref{fig:3DGS_TensorGS_overview}.

\textbf{(1) Tensorizing the power computation.}
We first reformulate the dominant Gaussian power calculation as a matrix-style computation so that many (Gaussian, pixel) interactions can be processed together via Tensor Core instructions, rather than being computed independently by scalar CUDA threads.


\textbf{(2) Grouping neighboring tiles for higher reuse and utilization.}
Direct tensorization alone is insufficient for high efficiency. In the original rasterization pipeline, each pixel tile is processed independently. However, neighboring tiles often overlap with many of the same Gaussians. As a result, the same Gaussian data must be fetched and processed repeatedly across nearby tiles, preventing effective reuse and often producing small per-tile workloads. To address this issue, TensorGS groups neighboring tiles into a larger processing region so that overlapping Gaussians can be reused across multiple tiles, reducing the cost of Gaussian fetching.

Together, these two components convert the original scalar rasterization hotspot into a Tensor-Core-friendly computation with substantially better arithmetic throughput and data reuse. We next describe these two components in detail.



\subsection{Tensorizing the Power Computation}
\label{sec:design_tensorizing}

We target the power computation, since it is the dominant arithmetic hotspot inside rasterization. 
To explain how this computation can be transformed into a Tensor-Core-friendly form, we begin from the formulation introduced in Section~\ref{sec:bg}, and then reorganize it into a batched matrix multiplication.

\subsubsection{Revisiting the Rasterization Calculation}
\label{sec:design_tensorizing_reformulate}
\hfill \break
\noindent To tensorize the 3DGS rasterization pipeline, we need to first dive deeper into its calculation. After feature computation, each visible Gaussian is projected onto the image plane, and is represented by a 2D mean
\begin{equation}
    \boldsymbol{\mu}'_g =
    \begin{bmatrix}
        \mu^x_g \\
        \mu^y_g
    \end{bmatrix},
\end{equation}
where $\mu^x_g$ and $\mu^y_g$ denote the horizontal and vertical coordinates of the projected Gaussian center on the image plane, respectively, together with a projected 2D covariance matrix
\begin{equation}
\label{eq:sigma}
    \boldsymbol{\Sigma}'_g =
    \begin{bmatrix}
        \sigma^{xx}_g & \sigma^{xy}_g \\
        \sigma^{xy}_g & \sigma^{yy}_g
    \end{bmatrix}.
\end{equation}
Here, $\boldsymbol{\Sigma}'_g$ describes the spatial spread of the projected Gaussian in screen space: $\sigma^{xx}_g$ and $\sigma^{yy}_g$ correspond to the variance along the horizontal and vertical directions, while $\sigma^{xy}_g$ captures the correlation between them. Geometrically, $\boldsymbol{\mu}'_g$ specifies the center of the Gaussian splat, and $\boldsymbol{\Sigma}'_g$ determines the size, shape, and orientation of its elliptical support on the image plane.

During rasterization, for a pixel
\begin{equation}
    \mathbf{x}_p =
    \begin{bmatrix}
        x_p \\
        y_p
    \end{bmatrix},
\end{equation}
where $x_p$ and $y_p$ denote the horizontal and vertical coordinates of pixel $p$ on the image plane, the renderer first computes its offset to the Gaussian center:
\begin{equation}
    \mathbf{d}_{g,p}
    =
    \mathbf{x}_p - \boldsymbol{\mu}'_g
    =
    \begin{bmatrix}
        x_p - \mu^x_g \\
        y_p - \mu^y_g
    \end{bmatrix}
    =
    \begin{bmatrix}
        \Delta x_{g,p} \\
        \Delta y_{g,p}
    \end{bmatrix}.
\end{equation}
This offset indicates where the current pixel lies relative to the Gaussian center in screen space. 
The power term is then calculated as
\begin{equation}
    \mathrm{power}_{g,p}
    =
    -\frac{1}{2}
    \mathbf{d}_{g,p}^{\top}
    (\boldsymbol{\Sigma}'_g)^{-1}
    \mathbf{d}_{g,p}.
\end{equation}
Here, $(\boldsymbol{\Sigma}'_g)^{-1}$ is the inverse of the projected covariance matrix $\boldsymbol{\Sigma}'_g$ in Equation~\ref{eq:sigma}, i.e., the screen-space precision matrix of Gaussian $g$. In rasterization, this inverse covariance is commonly referred to as the \emph{conic matrix}, since it defines the anisotropic quadratic form that measures the offset $\mathbf{d}_{g,p}$. To distinguish it clearly from the covariance entries in $\boldsymbol{\Sigma}'_g$, we denote it as
\begin{equation}
    (\boldsymbol{\Sigma}'_g)^{-1}
    =
    \begin{bmatrix}
        a_g & b_g \\
        b_g & c_g
    \end{bmatrix}.
\end{equation}
The three scalars $a_g$, $b_g$, and $c_g$ therefore fully characterize the Gaussian's conic form in screen space: $a_g$ and $c_g$ control the decay along the two coordinate directions, while $b_g$ captures the coupling between them.

Substituting the offset vector into the quadratic form gives
\begin{equation}
    \mathrm{power}_{g,p}
    =
    -\frac{1}{2}
    \begin{bmatrix}
        \Delta x_{g,p} & \Delta y_{g,p}
    \end{bmatrix}
    \begin{bmatrix}
        a_g & b_g \\
        b_g & c_g
    \end{bmatrix}
    \begin{bmatrix}
        \Delta x_{g,p} \\
        \Delta y_{g,p}
    \end{bmatrix},
\end{equation}
which can be expanded into the scalar form
\begin{equation}
    \mathrm{power}_{g,p}
    =
    -\frac{1}{2} a_g \Delta x_{g,p}^2
    - b_g \Delta x_{g,p}\Delta y_{g,p}
    -\frac{1}{2} c_g \Delta y_{g,p}^2.
    \label{eq:tensorgs_power_scalar}
\end{equation}
Equation~\ref{eq:tensorgs_power_scalar} shows that each power evaluation consists of two parts: three Gaussian-dependent coefficients,
\[
\left[
-\frac{1}{2}a_g,\;
-b_g,\;
-\frac{1}{2}c_g
\right],
\]
and three quadratic basis terms associated with the current Gaussian--pixel pair,
\[
\left[
\Delta x_{g,p}^2,\;
\Delta x_{g,p}\Delta y_{g,p},\;
\Delta y_{g,p}^2
\right]^{\top}.
\]
Therefore, although the baseline rasterizer evaluates the power term independently for each Gaussian--pixel pair using scalar CUDA operations, the computation itself is fundamentally a \emph{three-term reduction}. This observation is the starting point of our tensorization.

\subsubsection{From Scalar Calculation to Batched Computation}
\label{sec:design_tensorizing_matrix}
\hfill \break
In the original tile-based rasterizer, one thread block is assigned to one $16\times16$ tile, i.e., 256 pixels. 
Suppose we process a batch of $G_{\text{batch}}$ Gaussians against this tile. Then, instead of calculating Equation~\ref{eq:tensorgs_power_scalar} separately for each (Gaussian, pixel) pair, we organize all $G_{\text{batch}}\times256$ interactions into a batched matrix-style computation.

We first stack the Gaussian side coefficients into a matrix
\begin{equation}
    \mathbf{Q}
    =
    \begin{bmatrix}
        -\frac{1}{2}a_1 & -b_1 & -\frac{1}{2}c_1 \\
        -\frac{1}{2}a_2 & -b_2 & -\frac{1}{2}c_2 \\
        \vdots & \vdots & \vdots \\
        -\frac{1}{2}a_{G_{\text{batch}}} & -b_{G_{\text{batch}}} & -\frac{1}{2}c_{G_{\text{batch}}}
    \end{bmatrix}
    \in \mathbb{R}^{G_{\text{batch}}\times 3}.
\end{equation}
Each row corresponds to one Gaussian and stores the three conic coefficients required by Equation~\ref{eq:tensorgs_power_scalar}.

Next, for the current tile, we collect the three quadratic basis channels into $\mathbf{\Phi} \in \mathbb{R}^{3\times256},$
where each column corresponds to one pixel and contains the three basis terms 
$\left[
\Delta x^2,\;
\Delta x \Delta y,\;
\Delta y^2
\right]^{\top}
$
used in the power computation. 
For ease of exposition, we use $\mathbf{\Phi}$ to denote the three basis channels associated with the current Gaussian--tile workload; the key point is that the power computation reduces these three terms along the inner dimension.

With this organization, the power values of the current Gaussian batch over the current tile can be viewed as
\begin{equation}
    \mathbf{P} = \mathbf{Q}\mathbf{\Phi},
    \qquad
    \mathbf{P} \in \mathbb{R}^{G_{\text{batch}}\times256},
\end{equation}
where the output entry $\mathbf{P}_{g,p}$ corresponds to the power value of Gaussian $g$ at pixel $p$.

This reformulation exposes the two natural batching dimensions already present in rasterization: the Gaussian dimension and the pixel dimension. The baseline implementation traverses these pairs one by one in scalar form, whereas our reformulation groups them into a dense tile-level computation, thereby creating the opportunity to map the workload to Tensor Core matrix instructions.

\subsubsection{Padding the Inner Dimension for Tensor Core Execution}
\label{sec:design_tensorizing_padding}
\hfill \break
However, the matrix-style formulation above is still not directly compatible with Tensor Core instructions. The reason is that its inner dimension is only $K=3$, whereas NVIDIA WMMA instructions operate on fixed fragment shapes such as \texttt{m16n16k16}. In other words, although the power computation already exhibits a matrix-style structure, its native shape does not yet match the hardware granularity required by Tensor Cores.

To resolve this mismatch, we pad the inner dimension from 3 to 16:
\begin{equation}
    \widetilde{\mathbf{Q}} \in \mathbb{R}^{G_{\text{batch}} \times 16},
    \qquad
    \widetilde{\mathbf{\Phi}} \in \mathbb{R}^{16 \times 256},
\end{equation}
where the first three channels store the original values and the remaining 13 channels are filled with zeros:
\begin{equation}
    \widetilde{\mathbf{Q}}_{g,k}
    =
    \begin{cases}
        \mathbf{Q}_{g,k}, & k \le 3, \\
        0, & 4 \le k \le 16,
    \end{cases}
    \qquad
    \widetilde{\mathbf{\Phi}}_{k,p}
    =
    \begin{cases}
        \mathbf{\Phi}_{k,p}, & k \le 3, \\
        0, & 4 \le k \le 16.
    \end{cases}
\end{equation}
The padded multiplication is
\begin{equation}
    \widetilde{\mathbf{P}}
    =
    \widetilde{\mathbf{Q}}
    \widetilde{\mathbf{\Phi}}
    \in \mathbb{R}^{G_{\text{batch}} \times 256}.
\end{equation}

Importantly, this padding introduces \emph{no numerical error}. For any Gaussian $g$ and pixel $p$,
\begin{equation}
    \widetilde{\mathbf{P}}_{g,p}
    =
    \sum_{k=1}^{16}
    \widetilde{\mathbf{Q}}_{g,k}
    \widetilde{\mathbf{\Phi}}_{k,p}
    =
    \sum_{k=1}^{3}
    \mathbf{Q}_{g,k}\mathbf{\Phi}_{k,p}
    +
    \sum_{k=4}^{16}
    0 \cdot 0
    =
    \mathbf{P}_{g,p}.
\end{equation}
That is, the additional padded channels are mathematically inactive: they exist only to match the fragment shape required by Tensor Core instructions but do not change the computed power values.

\subsubsection{Warp-Level WMMA Mapping and Operand Layout}
\label{sec:design_tensorizing_wmma}
\hfill \break
In addition to the padded 16-channel inner dimension, the Gaussian dimension and the pixel dimension are also blocked into groups of 16 to match the native \texttt{m16n16k16} WMMA shape. As a result, each WMMA operation produces a $16\times16$ output block of power values corresponding to \textbf{16 Gaussians} and \textbf{16 pixels}: the 16 rows index 16 Gaussians, the 16 columns index 16 pixels, and each output entry stores the power value of one Gaussian--pixel pair. Accordingly, one warp computes this block as: $\mathbf{C}_{16\times16} = \mathbf{A}_{16\times16} \mathbf{B}_{16\times16}$, where $\mathbf{C}$ is the output matrix, $\mathbf{A}$ is the Gaussian-side operand, and $\mathbf{B}$ is the pixel-side operand. More specifically, each row of $\mathbf{A}$ stores the padded conic coefficients of one Gaussian,
\[
\left[-\frac{1}{2}a_g,\,-b_g,\,-\frac{1}{2}c_g,\;0,\dots,0\right],
\]
while each column of $\mathbf{B}$ stores the padded quadratic basis terms of one pixel,
\[
\left[\Delta x^2,\;\Delta x\Delta y,\;\Delta y^2,\;0,\dots,0\right]^{\top}.
\]
Thus, the $16\times16$ shape of $\mathbf{C}$ refers to the output block size, whereas the padded 16 in $\mathbf{A}$ and $\mathbf{B}$ refers to the inner reduction dimension.

At the thread-block level, a \emph{cooperative thread array} (CTA) iterates over the tile in 16-pixel panels along the column dimension of the output. Since one tile contains 256 pixels, the full tile is decomposed into 256 / 16 = 16 panels. If $G_{\text{batch}} > 16$, the CTA also iterates over the Gaussian dimension in chunks of 16. Hence, the complete tile computation is assembled from a sequence of warp-level WMMA operations, each of which calculates one $16$-Gaussian $\times$ $16$-pixel output block.

To feed these WMMA operations efficiently, both operands are loaded into shared memory before being loaded into WMMA fragments. The Gaussian-side operand is reused across multiple 16-pixel panels within the same tile, while the pixel-side operand is prepared panel by panel in the padded fragment layout required by WMMA.
This design provides two benefits. First, the Gaussian coefficients are loaded from global memory only once per batch and then reused from shared memory across multiple pixel panels. Second, storing both operands directly in the padded WMMA-compatible layout avoids extra data reorganization and control flow inside the warp.

\subsection{Limitations of Naive Tensorization}
\label{sec:design_limitation}

\begin{table}[t]
    \centering
    \small
    \setlength{\tabcolsep}{3pt}
    \caption{Naive Tensorization vs. original 3DGS. The hardware utilization metrics are measured during the rasterization stage, and the results are averaged over all six scenes.}
    \label{tab:naive_tensorgs_vs_baseline}
    \begin{tabular}{p{2.55cm} m{1.2cm}<{\centering} m{1.5cm}<{\centering} m{2.1cm}<{\centering}}
        \toprule
        Method & Speedup & Tensor Core Util. & DRAM Bandwidth Util. \\
        \midrule
        Original 3DGS       & 1.00$\times$ & 0\%  & 38\% \\
        Naive Tensorization & 1.29$\times$ & 31\% & 76\% \\
        \bottomrule
    \end{tabular}
\end{table}

The tensorization in Section~\ref{sec:design_tensorizing} successfully maps the power computation onto Tensor Core instructions. However, this algebraic reformulation alone is not sufficient to deliver a large end-to-end speedup. In particular, a naive implementation that applies the tensorized computation independently to each tile yields only limited improvement over the original 3DGS rasterizer.

Table~\ref{tab:naive_tensorgs_vs_baseline} compares the naive tensorized version against the original 3DGS implementation. As shown in the table, naive tensorization effectively activates Tensor Cores, but the achieved speedup remains modest at only 1.29$\times$. Meanwhile, Tensor Core utilization increases from 0\% in the baseline to 31\%, while DRAM bandwidth utilization rises substantially from 38\% to 76\%.

These results suggest that the bottleneck shifts after naive tensorization. Although Tensor Cores accelerate the arithmetic part of the power computation, the overall kernel becomes increasingly constrained by data movement, so the additional arithmetic capability of Tensor Cores cannot be translated into a significant end-to-end speedup. Once the arithmetic part is accelerated, data loading and operand feeding become the bottlenecks, and the kernel cannot keep Tensor Cores sufficiently busy.
In other words, the limitation of naive tensorization is not whether the power computation can run on Tensor Cores, but whether the surrounding workload organization can supply them efficiently enough to sustain high Tensor Core utilization. Simply enabling WMMA instructions is therefore insufficient to deliver a large end-to-end speedup.





\subsection{Cross-Tile Grouping for Gaussian Reuse}
\label{sec:design_grouping}

\subsubsection{Observation: Neighboring Tiles Share Many Gaussians}
\label{sec:design_grouping_observation}
\hfill \break
The limited speedup and utilization of naive tensorization suggest that the per-tile execution granularity is still inefficient for Tensor Core execution.
To identify a better execution granularity, we revisit how rasterization is organized in the original 3DGS pipeline.
There, each Gaussian is instantiated according to the tiles it overlaps, and each tile is processed independently by a separate CTA (i.e., thread block).
As a result, even if neighboring tiles share many overlapping Gaussians, the same Gaussian data will still be repeatedly loaded and processed across multiple tiles.

However, the projected Gaussians indeed span multiple neighboring tiles in screen space~\cite{10.1145/3592433}.
Consequently, neighboring tiles tend to have highly overlapping Gaussian lists, creating a natural opportunity for cross-tile reuse.

To quantify this effect, we measure how much Gaussian loading can be reduced by processing neighboring tiles together. In this experiment, we set the group size to $2\times2$ tiles. For a tile group, let $N_{\text{total}}$ denote the total number of Gaussian appearances across all four tiles when they are processed independently, and let $N_{\text{group}}$ denote the number of Gaussian appearances after merging the four tiles into one group. We define the Gaussian loading reduction ratio as: $\text{LoadReduction} = 1-N_{\text{group}} / N_{\text{total}}$.

\begin{figure}[t]
    \centering
    \includegraphics[width=0.98\linewidth]{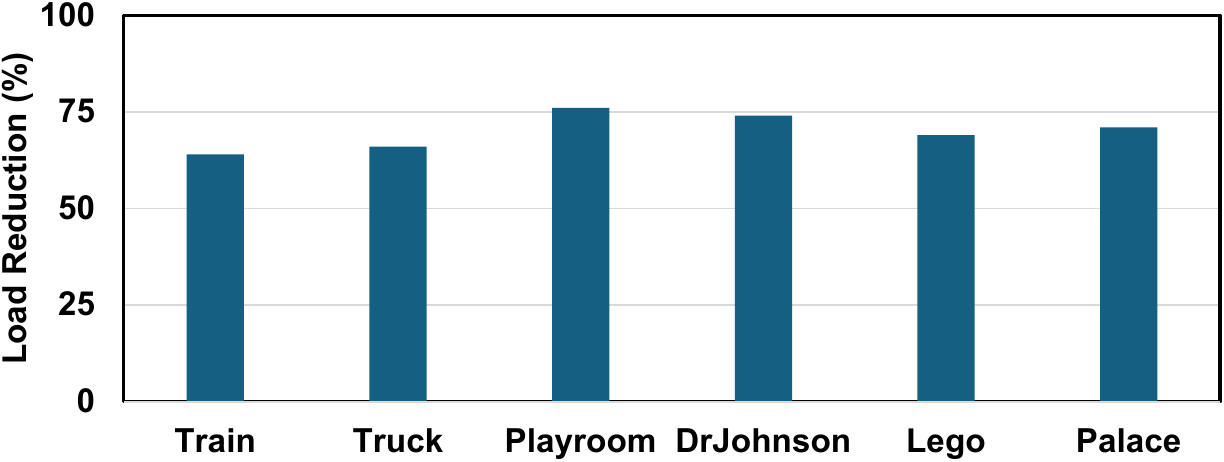}
    \caption{Gaussian loading reduction within the default $2\times2$ tile group across six scenes.
    }
    \label{fig:tile_group_reuse}
\end{figure}

As shown in Figure~\ref{fig:tile_group_reuse}, the reduction is substantial for a $2\times2$ tile group. Averaged over all scenes, the Gaussian loading reduction reaches 70\%, corresponding to a reuse factor of more than 3$\times$. 
In other words, within a $2\times2$ tile group, one Gaussian overlaps more than three tiles on average. However, with the baseline independent tile processing, we would still load the Gaussian separately for each tile.
To this end, cross-tile grouping is a practical way to improve Gaussian reuse.



\subsubsection{Grouped Rasterization with Shared Gaussian}
\label{sec:design_grouping_impl}
\hfill \break
\textbf{Co-locating neighboring tiles on the same SM.}
The goal of our design is to place neighboring tiles on the same SM so that they can share SM-local resources, especially shared memory. CUDA does not provide explicit control over block-to-SM assignment, so we cannot directly pin selected tiles to a specific SM. However, CUDA does guarantee that a \emph{cooperative thread array} (CTA), i.e., a thread block, is executed on a single SM and is never split across multiple SMs. We therefore achieve co-location indirectly by assigning a group of neighboring tiles to one CTA. Once grouped into the same CTA, these tiles are guaranteed to execute on the same SM and can reuse Gaussian data loaded in shared memory.

Importantly, we do \emph{not} enlarge the CTA by assigning one thread to every pixel of every tile in the group, since that strategy would quickly exceed the maximum number of threads per block for large group sizes. Instead, we keep the CTA size fixed and let the CTA iterate over the tiles inside the group. In this way, the tile group becomes the logical execution unit for reuse, while the CTA remains a valid CUDA execution unit.

\textbf{Constructing the grouped Gaussian list.}
We partition the image into tile groups of size $H_g \times W_g$, where each group contains $H_g \times W_g$ non-overlapping neighboring tiles. TensorGS uses a $2\times2$ tile group as the default configuration, and we conduct a sensitivity study of other group sizes in Section~\ref{sec:eval_sensitivity}.
The grouped Gaussian list is constructed before rasterization, during the Gaussian--tile intersection test in the feature computation stage of the standard 3DGS pipeline (Section~\ref{sec:bg}). Specifically, for each projected Gaussian, we first identify the set of overlapped tiles according to its screen-space footprint, and then map these tiles to their corresponding tile groups. In the baseline implementation, one duplicated entry is generated for every overlapped tile. In contrast, TensorGS generates one entry for every overlapped \emph{group}. If a Gaussian overlaps multiple tiles within the same group, these duplicated tile-level entries are merged into a single group-level entry.

Each group-level entry stores not only the Gaussian identifier and its depth, but also a compact \emph{tile-membership bit mask} indicating which member tiles in the group are actually overlapped by that Gaussian. For a group of size $H_g \times W_g$, the mask contains $H_g W_g$ bits. For example, a $2\times2$ group requires only 4 bits, while a $4\times4$ group requires 16 bits. Using this encoding, all Gaussian entries belonging to the same group are then sorted by \emph{(group ID, depth)}, producing a depth-ordered Gaussian list for each tile group before rasterization begins.
We provide a detailed overhead analysis in Section \ref{sec:eval_overhead}.

\textbf{Grouped rasterization.}
During rasterization, one \emph{cooperative thread array} (CTA), i.e., one thread block, is assigned to one tile group and processes the corresponding depth-sorted group-level Gaussian list chunk by chunk. Importantly, the entire group-level list is \emph{not} loaded into shared memory at once. Instead, at each step, only a small Gaussian chunk is loaded into shared memory and transformed into the Gaussian-side WMMA operand layout. The same loaded chunk is then reused while the CTA iterates over the tiles in the group. After the current chunk is consumed, the CTA proceeds to the next chunk in the list. In this way, cross-tile grouping increases Gaussian reuse without the need to fit the entire group-level list into shared memory.

Within the CTA, each tile still maintains its own blending state, including transmittance, color accumulation, and early-termination condition. Therefore, grouping tiles does not change the rendering semantics. It only changes the granularity at which Gaussian data is loaded and reused. 
For each loaded Gaussian entry, a tile first checks the tile-membership bit mask to determine whether that Gaussian overlaps the tile. 
Only if the corresponding bit is set does the tile use that Gaussian to construct its tile-specific pixel-side WMMA operands and send them to Tensor Cores for computation. Otherwise, the Gaussian is skipped for that tile. Since the group-level list is already sorted by depth, filtering it with the tile-membership bit mask preserves the original front-to-back order for every tile. As a result, the grouped rasterizer remains numerically equivalent to the original tile-wise rasterization, while amortizing Gaussian loading and tensor preparation across multiple neighboring tiles.


\subsection{Overall Workflow}
\label{sec:design_workflow}

Algorithm~\ref{alg:tensorgs} shows the workflow of TensorGS. Similar to the standard 3DGS renderer, TensorGS first performs frustum culling and feature computation. It then constructs depth-sorted group-level Gaussian lists with tile-membership bit masks. During rasterization, each cooperative thread array (CTA) processes one tile group. For each tile in the group, the tile-membership bit mask is used to determine which Gaussian entries are actually relevant to that tile. Only these relevant Gaussian entries are then used to construct the tile-specific WMMA operands for power evaluation on Tensor Cores. The remaining steps, including opacity computation, alpha blending, transmittance update, early termination, and final color accumulation, remain on CUDA cores.

\begin{algorithm}[t]
\caption{TensorGS workflow}
\label{alg:tensorgs}
\begin{algorithmic}[1]
\Require Gaussians $\mathcal{G}$, camera view $\mathcal{V}$, tile-group size $(H_g,W_g)$
\Ensure Rendered image $\mathbf{I}$

\State Frustum-cull visible Gaussians under $\mathcal{V}$
\State Partition the tile grid into $H_g \times W_g$ tile groups
\ForAll{visible Gaussian $g$}
    \State Compute screen-space attributes of $g$
    \State Find the tiles overlapped by $g$
    \ForAll{tile groups $u$ intersected by $g$}
        \State Generate one group-level Gaussian entry for $(g,u)$ together with a tile-membership bit mask
    \EndFor
\EndFor
\State Sort Gaussian entries by \emph{(group ID, depth)}

\ForAll{tile groups $u$ \textbf{in parallel}}
    \State Assign one CTA to $u$ and initialize per-tile blending states
    \ForAll{Gaussian chunks $\mathcal{B}$ in the sorted list of $u$}
        \State Load Gaussian data of $\mathcal{B}$ into shared memory
        \ForAll{tiles $t$ in group $u$}
            \State Select Gaussian entries in $\mathcal{B}$ whose bit mask includes $t$
            \State Power computation on Tensor Cores
            \State Compute opacity, alpha blending, transmittance update, early termination, and color accumulation on CUDA cores
        \EndFor
    \EndFor
\EndFor

\State \Return $\mathbf{I}$
\end{algorithmic}
\end{algorithm}




\section{Evaluation}
\label{sec:eval}

\subsection{Experiment Setup}
\label{sec:eval_setup}

\textbf{Platform, datasets, and scenes.}
We conduct all experiments on a server equipped with an NVIDIA A100 GPU (40GB).
We evaluate on six representative scenes from four datasets: Tanks\&Temples~\cite{knapitsch2017tanks}, Deep Blending~\cite{hedman2018deep}, Synthetic-NeRF~\cite{mildenhall2021nerf}, and Synthetic-NSVF~\cite{liu2020neural}.
Table~\ref{tab:eval_workloads} summarizes the evaluated workloads.

\begin{table}[t]
    \centering
    \caption{Evaluated workloads.}
    \label{tab:eval_workloads}
    \setlength{\tabcolsep}{3.5pt}
    \small
    \begin{tabular}{l l l}
        \toprule
        \textbf{Dataset} & \textbf{Scene (Resolution)} & \textbf{Type} \\
        \midrule
        \multirow{2}{*}{Tanks\&Temples}
            & Train ($980 \times 545$) & \multirow{2}{*}{Real-world outdoor} \\
            & Truck ($979 \times 546$) & \\
        \midrule
        \multirow{2}{*}{Deep Blending}
            & Playroom ($1264 \times 832$) & \multirow{2}{*}{Real-world indoor} \\
            & DrJohnson ($1332 \times 876$) & \\
        \midrule
        Synthetic-NeRF
            & Lego ($800 \times 800$) & Synthetic \\
        Synthetic-NSVF
            & Palace ($800 \times 800$) & Synthetic \\
        \bottomrule
    \end{tabular}
\end{table}

\textbf{Compared methods and implementation.}
We first compare our proposed TensorGS with the original 3DGS pipeline~\cite{10.1145/3592433} and our internal Naive Tensorization method.
We then select several representative 3DGS optimization methods, including gsplat~\cite{ye2025gsplat}, AdR-Gaussian~\cite{wang2024adr}, FlashGS~\cite{feng2025flashgs}, and GSCore~\cite{lee2024gscore}.
For each optimization method, we measure the speedup when integrated with our proposed TensorGS.
These methods represent different optimization directions, including optimized CUDA implementation, tighter intersection testing via refined bounding boxes, early culling and load balancing, and kernel-level rasterization optimization.
For methods involving hardware-specific support, we evaluate only their software components on the A100 GPU.

All methods are evaluated using the same pretrained Gaussian models and the same rendering configuration under a unified FP16 setting.
The resulting image quality remains nearly unchanged compared to FP32, with detailed quality results reported in Table~\ref{tab:moti_fp16_quality} and Section~\ref{sec:moti}.
Unless otherwise stated, TensorGS uses a default tile-group size of \textbf{$2 \times 2$} tiles. 
We further analyze this design choice and study its sensitivity in Section~\ref{sec:eval_sensitivity}.
All the results are averaged over 100 rendered frames.

\subsection{Main Results}
\label{sec:eval_main}

\begin{figure}[t]
    \centering
    \includegraphics[width=1\linewidth]{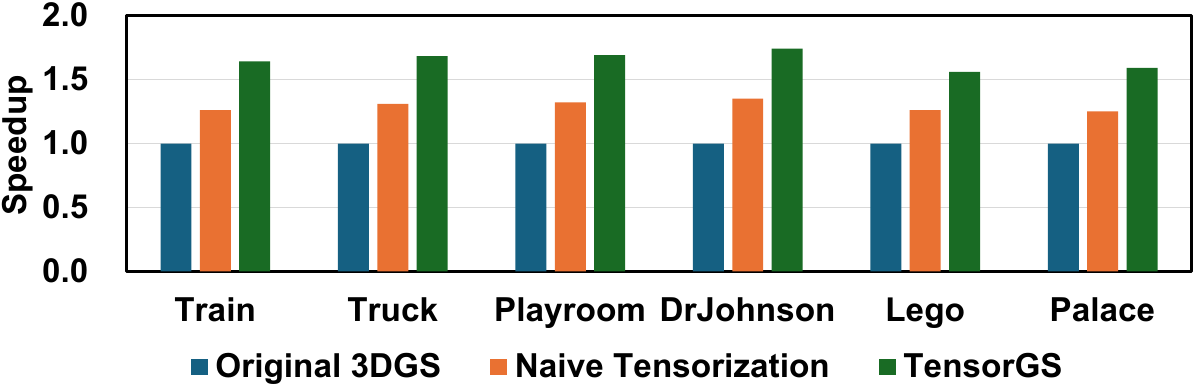}
    \caption{Speedup over the original 3DGS pipeline on six representative scenes.}
    \label{fig:eval_main_compare}
\end{figure}

\begin{figure}[t]
    \centering
    \includegraphics[width=1\linewidth]{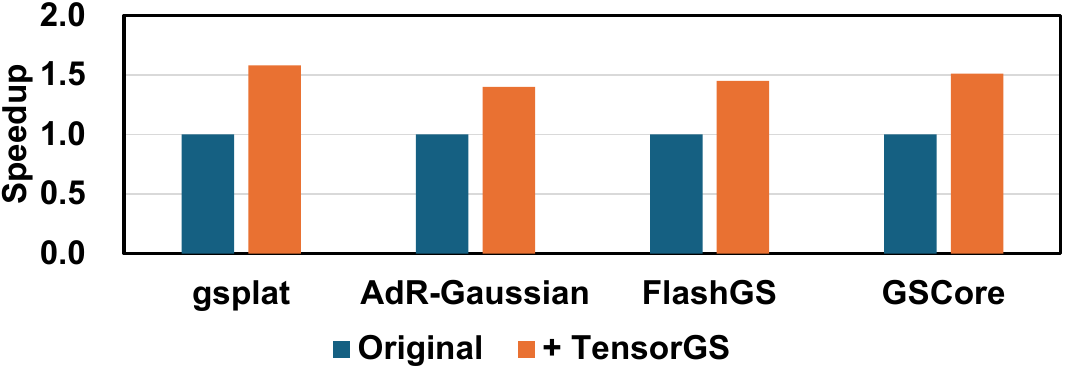}
    \caption{Speedup (averaged over six scenes) when integrating TensorGS into the four representative 3DGS optimization methods.}
    \label{fig:eval_main_integration}
\end{figure}

Figures~\ref{fig:eval_main_compare} and~\ref{fig:eval_main_integration} show the main evaluation results from two perspectives.
First, Figure~\ref{fig:eval_main_compare} compares Naive Tensorization and TensorGS against the original 3DGS pipeline on six representative scenes.
Second, Figure~\ref{fig:eval_main_integration} reports the end-to-end improvement obtained by integrating TensorGS into several representative 3DGS optimization methods, including gsplat, AdR-Gaussian, FlashGS, and GSCore.

Figure~\ref{fig:eval_main_compare} shows that naive tensorization alone brings only limited benefit.
Compared with the original 3DGS pipeline, Naive Tensorization achieves an average speedup of 1.29$\times$.
Although it offloads the power computation to Tensor Cores, it does not change the tile processing granularity, and therefore cannot expose sufficient Gaussian reuse.
By contrast, TensorGS achieves an average speedup of 1.65$\times$ over the original 3DGS pipeline.
Moreover, we also measure the Tensor Core utilization and find it increases from 31\% under Naive Tensorization to about 52\% under TensorGS, indicating that grouped execution provides a much more effective method for tensorized rasterization.
These results show that the grouping-based reorganization for tile processing is necessary to translate Tensor-Core support into more substantial end-to-end speedup.

The gain of TensorGS is generally more pronounced on scenes with heavier rasterization workloads (e.g., DrJohnson scene), where more Gaussian--tile interactions must be processed, and the opportunity for cross-tile reuse is greater.
On lighter scenes (e.g., the two synthetic scenes), the improvement is somewhat smaller but still noticeable, indicating that TensorGS is robust across different scenes.

Figure~\ref{fig:eval_main_integration} further shows that TensorGS consistently improves multiple representative 3DGS optimization methods.
On top of gsplat, AdR-Gaussian, FlashGS, and GSCore, TensorGS brings additional speedups of 1.58$\times$, 1.40$\times$, 1.45$\times$, and 1.51$\times$, respectively.
These results indicate that TensorGS is largely complementary to existing optimization directions, consistently delivering additional end-to-end gains on top of already optimized rasterization methods.

\subsection{Overhead Analysis}
\label{sec:eval_overhead}

The overhead in TensorGS mainly comes from two sources: i) the tile-membership bit mask attached to each Gaussian and ii) the construction of group-level Gaussian lists in the grouped rasterization pipeline.

\textbf{Bit-mask storage overhead.}
For the tile group, each Gaussian only needs to record whether it belongs to each of the member tiles in the group.
For example, in our default $2\times2$ group size configuration, the tile-membership mask requires only 4 bits.
In our implementation, we store this mask using one \texttt{uint8\_t}, i.e., one byte.
On the other hand, each Gaussian entry in the grouped rasterization pipeline consists of the projected 2D center position, conic-opacity parameters, and a Gaussian ID.
Under FP16, these fields occupy 16 bytes in total: 4 bytes for the 2D center position, 8 bytes for the conic-opacity parameters, and 4 bytes for the Gaussian ID.
As a result, the bit mask adds only 1 extra byte over the original 16-byte Gaussian entry, corresponding to \textbf{6.25\%} additional storage overhead.


Although the bit mask slightly increases the per-Gaussian metadata size, it enables a single Gaussian entry to be shared across multiple neighboring tiles within the same tile group.
Without such a mask, the same Gaussian would need to be duplicated in multiple per-tile lists.
Therefore, the added one-byte overhead is a compact price for encoding tile membership, while reducing redundant Gaussian duplication and enabling cross-tile reuse.

\textbf{Runtime overhead.}
In addition to the storage overhead above, TensorGS introduces extra runtime cost from building the group-level Gaussian lists and evaluating the tile-membership mask during rasterization.
Specifically, compared with the original per-tile organization, TensorGS needs to i) generate the membership mask when forming the grouped list, and ii) perform a lightweight bit test before processing a tile--Gaussian pair.
Fortunately, these operations consist only of simple integer and bitwise logic, and their cost is small compared with the arithmetic-heavy Gaussian evaluation accelerated by Tensor Cores.
Importantly, all the reported speedups have already included this overhead.


\subsection{Analysis and Sensitivity Study on Tile Grouping}
\label{sec:eval_sensitivity}

A key design parameter in TensorGS is tile grouping.
Using smaller tile groups limits the opportunity to reuse Gaussian data across neighboring tiles, while using larger tile groups enlarges the group-level Gaussian list and introduces more Gaussians that intersect only a small subset of tiles within each group.
As a result, more non-contributing (Gaussian, tile) pairs must be filtered by the tile-membership mask.
The group size therefore, determines the trade-off between cross-tile reuse and grouped-rasterization overhead.

In TensorGS, we group neighboring tiles into square $G\times G$ blocks and evaluate $G\in\{1,2,4,8\}$.
The $1\times1$ case reduces to the original per-tile organization and therefore serves as a reference point with no cross-tile reuse.
We use square groups because each projected Gaussian spans a two-dimensional region in screen space, so tile grouping should naturally account for both horizontal and vertical overlap rather than only one-dimensional adjacency.


\begin{table}[t]
    \centering
    \caption{Sensitivity study on tile-group size. The results are the average over six scenes.}
    \label{tab:eval_group_size}
    \setlength{\tabcolsep}{4pt}
    \begin{tabular}{m{2.2cm}<{\centering} m{1.1cm}<{\centering} m{1.1cm}<{\centering} m{1.1cm}<{\centering} m{1.1cm}<{\centering}}
        \toprule
        Group Size & $1\times1$ & $2\times2$ & $4\times4$ & $8\times8$ \\
        \midrule
        Speedup over standard 3DGS  & 1.29$\times$ & $1.65\times$ & 1.56$\times$ & 1.43$\times$ \\
        \bottomrule
    \end{tabular}
\end{table}

Table~\ref{tab:eval_group_size} shows the sensitivity to tile-group size.
When there is no grouping (i.e., $1\times1$ tile), TensorGS cannot exploit reuse across neighboring tiles, so the benefit of tensorizing the computation remains limited.
As the group size increases from $1\times1$ to $2\times2$, performance improves because Gaussian data can be shared within each group.
However, further increasing the group size leads to diminishing returns and eventually degrades performance.
This is because larger groups produce longer grouped Gaussian lists and include more Gaussians that contribute to only a small subset of tiles within each group.
As such, we use the $2\times2$ group size as the default setting in our design.

\section{Related Works}
\label{sec:related}

A number of recent works have explored accelerating 3D Gaussian Splatting from different angles, spanning algorithmic optimizations, dedicated hardware accelerators, and training-side improvements.

Some works focus on reducing redundant computation early in the rendering pipeline. GSCore~\cite{lee2024gscore} observes that the vanilla AABB-based intersection test can produce many false-positive Gaussian--tile pairs for anisotropic splats, and adopts a tighter shape-aware support estimation together with hierarchical sorting and subtile skipping in a dedicated accelerator. FlashGS~\cite{feng2025flashgs} further improves this stage with an opacity-aware precise ellipse--tile intersection test on commodity GPUs, showing that more faithful Gaussian support estimation can substantially reduce the downstream cost of sorting and rasterization. GCC~\cite{pei2025gcc} is also related in spirit, as it introduces alpha-based boundary identification and cross-stage conditional processing to derive more compact effective Gaussian regions and avoid preprocessing or rendering ineffective Gaussians.

Other works target later stages of the rendering pipeline. Neo~\cite{oh2026neo} focuses on the sorting stage and exploits temporal redundancy across consecutive frames through reuse-and-update sorting rather than re-sorting from scratch. There are also works that target the blending stage. GBU~\cite{ye2025gaussian} reduces blending cost through adjacent-pixel computation sharing and fragment skipping on edge GPUs, while ORANGE~\cite{li2026orange} reformulates 3DGS blending into a GEMM-friendly form for DNN-oriented NPUs. These efforts are largely complementary to preprocessing-side optimizations, as they focus more on reducing the cost of ordering or blending once Gaussian--tile assignments have already been formed.

There are also works that emphasize training efficiency or memory scalability. GSArch~\cite{he2025gsarch} and Cambricon-GS~\cite{wen2026cambricon} accelerate training by addressing memory barriers or invalid $\alpha$ computation during forward and backward propagation, whereas CLM~\cite{zhao2026clm} and GS-Scale~\cite{lee2026gs} scale 3DGS training to larger scenes via host offloading. In addition, some recent efforts focus specifically on 3DGS-SLAM rather than general-purpose novel-view synthesis. Examples include GauSPU~\cite{wu2024gauspu}, RTGS~\cite{li2025rtgs}, SPLATONIC~\cite{huang2026splatonic}, REACT3D~\cite{wang2025react3d}, and AGS~\cite{he2026ags}, which exploit task-specific sparsity, redundancy reduction, or temporal coherence in SLAM tracking and mapping.

Compared with these prior efforts, our work differs in both optimization target and mechanism. Prior methods such as GSCore, FlashGS, and GCC mainly reduce downstream cost by tightening Gaussian support regions or skipping ineffective Gaussian--tile pairs. Neo focuses on sorting reuse across frames, while GBU and ORANGE focus on blending-side acceleration through adjacent-pixel computation sharing or GEMM-friendly execution on alternative hardware backends. In contrast, our work targets Tensor-Core acceleration of the volume-rendering computation itself, aiming to improve the efficiency of the core rendering computation rather than optimizing preprocessing, sorting, or task-specific SLAM/training pipelines.

Compared with these prior efforts, our work has a different focus. While many existing methods emphasize tighter intersection, sorting reuse, or system-level acceleration, our work focuses on improving the efficiency of the volume-rendering computation itself through Tensor-Core acceleration. In this sense, our work is largely complementary to prior efforts.

\section{Conclusion}
\label{sec:conclusion}

In this paper, we present TensorGS, a Tensor-Core-accelerated 3D Gaussian Splatting framework that improves the efficiency of rasterization. TensorGS reformulates the dominant power computation into Tensor-Core-friendly matrix operations and introduces cross-tile grouping to improve Gaussian reuse across neighboring tiles. Experimental results show that TensorGS achieves 1.65$\times$ end-to-end speedup compared to the standard 3DGS pipeline while preserving rendering quality. Moreover, TensorGS remains complementary to many existing 3DGS optimization methods, consistently delivering additional performance gains when integrated with them. 
Overall, these results show that Tensor-Core-aware reformulation of 3DGS rasterization can deliver substantial end-to-end rendering acceleration on modern GPUs.


\bibliographystyle{ACM-Reference-Format}
\bibliography{main}










\end{document}